\documentstyle[11pt,twoside,jltp]{article}

\title{Momentum-Space Topology of
Standard Model}

\author{G.E. Volovik\address{Low Temperature Laboratory,
Helsinki University of Technology,
P.O.Box 2200, FIN-02015 HUT, Finland\\
L.D. Landau Institute for
Theoretical Physics,  Kosygin Str. 2, 117940 Moscow, Russia}}

\runninghead{G.E. Volovik}{Momentum-Space Topology of
Standard Model}

\begin{document}

\begin{abstract}
The momentum-space topological invariants, which characterize the ground
state of the Standard Model, are continuous functions of two
parameters, generated by the hypercharge and by the weak charge. These
invariants provide the absence of the mass of the elementary fermionic
particles in the symmetric phase above the electroweak transition (the mass
protection). All the invariants become zero in the broken symmetry phase, as a
result all the elementary fermions become massive. Relation of the
momentum-space invariants to chiral anomaly is also discussed.

\end{abstract}

\maketitle


\section{INTRODUCTION}
There is an important class of the 3+1 dimensional Fermi systems. The  systems
which belong to this class are characterized by the Fermi points instead of the
Fermi surfaces. Among the representatives of this class are the vacuum of
Standard Model of  electroweak and strong interactions in high energy physics
and the superfluid
$^3$He-A in condensed matter physics  \cite{PNAS}.   Fermi point is the point
in the momentum space ${\bf p}$ where the (quasi)particle energy is zero. In
particle physics the energy spectrum $E({\bf p})=cp$ is characteristic of the
massless neutrino (or any other chiral lepton or quark in the Standard Model)
with $c$ being the speed of light. The energy of the chiral neutrino is zero at
point
${\bf p}=0$ in the 3D momentum space. In condensed matter systems the points
with zero energy have been realized first in superfluid $^3$He-A, where the
superfluid gap has two point nodes.  Another example of the
Fermi point in condensed matter system has been discussed for gapless
semiconductors
\cite{Abrikosov}.

The massless (gapless) character of the
fermionic spectrum in the system with the Fermi point is protected by the
topological invariant of the ground state, which is expressed as the integral
over the Green's function in the 4D momentum-frequency space:
\begin{equation}
N = {\bf tr} ~{\cal N}~~,~~{\cal N}  =
{1\over{24\pi^2}}e_{\mu\nu\lambda\gamma}~
 ~\int_{\sigma}~  dS^{\gamma}
~ {\cal G}\partial_{p_\mu} {\cal G}^{-1}
{\cal G}\partial_{p_\nu} {\cal G}^{-1} {\cal G}\partial_{p_\lambda}  {\cal
G}^{-1}~.
\label{TopInvariantMatrix}
\end{equation}
The integral here is over the surface
$\sigma$ embracing the point ${\bf p}=0, p_0=0$.
in the ¤4D momentum space ${\bf p}, p_0$, and $p_0$ is the energy (frequency)
along the imaginary axis; ${\bf tr}$ is the trace over the fermionic indices.

As an example let us consider the chiral spin-1/2 particle. In the limit of the
noninteracting particles, the Green's function matrix considered on the
imaginary
frequency axis,
$z=ip_0$, is
\begin{equation}
{\cal G} =(ip_0-{\cal H})^{-1}~~,~~{\cal H}=\pm  c\vec \sigma\cdot {\bf p}~.
\label{GreenFunction}
\end{equation}
Here    $\vec \sigma$ are $2\times 2$   Pauli spin matrices, so that
${\bf tr}$ in Eq.(\ref{TopInvariantMatrix}) is the trace over the spin
indices; the sign
$+$ is for a right-handed particle  and $-$  for a left-handed one; the spin of
the particle is oriented  along or opposite to its momentum, respectively. The
Green's function has a singularity at the point ${\bf p}=0, p_0=0$, and such
singularity cannot be removed because of the momentum-space topology.
Substituting this Green's function into the topological invariant,
Eq.(\ref{TopInvariantMatrix}), one obtains that this invariant is nonzero:
it is
$N=1$ for the righthanded particle and $N=-1$ for the lefthanded one.

What is important here that the Eq.(\ref{TopInvariantMatrix}), being the
topological invariant, does not change under any (but not very large)
perturbations. This means that even if the interaction between the particles
is introduced and the Green's functions changes drastically, the result remains
the same: $N=1$ for the righthanded particle and $N=-1$ for the lefthanded one.
The singularity of the Green's function remains, which means that the
quasiparticle spectrum remain gapless: fermions are massless even in the
presence of interaction.   The nonzero value of the momentum-space topological
invariant thus provides the mass protection for fermions. This mass protection
mechanism  based on the topological properties in the momentum space  is
ideologically  different from that based on gauge invariance
arguments \cite{FroggattNielsen}.

Now let us consider what happens if there are several fermionic species. In
this
case the trace operation is also over all the fermions. If the fermionic system
has an equal number of lefthanded and righthanded fermions, then the
topological
invariant in  Eq.(\ref{TopInvariantMatrix}) is $N=0$, and the above mass
protection mechanism does not work. This situation occurs in the planar phase
of superfluid $^3$He  (for planar phase see \cite{Vollhardt}) and the Standard
Model. If the righthanded neutrino is present, as follows from the Kamiokande
experiments, then each generation contains 8 lefthanded and 8 righthanded
fermions.

Here we show that, as in the case of the planar state of superfluid $^3$He,
there are modified topological invariants which provide the mass protection
even
for equal number of left and right fermions.

\section{Generating Function for Invariants}

One can introduce these invariants in the following way:
\begin{equation}
N(F) = {\bf tr} \left[{\cal N F}\right]~,
\label{TopInvariant}
\end{equation}
where ${\cal F}$ is any matrix function of any operator which commutes with the
Green's function ${\cal G}$. For example, in the Standard Model above the
electroweak transition there is a $U(1)_Y$ symmetry generated by the
hypercharge
$Y$. Since the hypercharge matrix ${\cal Y}$ commutes with the Green's
function,
the matrix
${\cal F}$ can be any  power of the hypercharge,
${\cal F}={\cal Y}^n$. One can easily verify that the perturbations of the
Green's function, which conserve the $U(1)_Y$ symmetry, do not change the
integral $N(Y^n)$. In the planar phase of $^3$He the corresponding $U(1)$
symmetry is combined rotations in spin and orbital space with generator
$J_3=S_3+L_3$.

Let us introduce the generating function for all the topological
invariants containing powers of the hypercharge
\begin{equation}
N(\theta_Y) ={\bf tr}  \left[ e^{i\theta_Y {\cal Y} } {\cal N} \right]~.
\label{GeneratingFunction}
\end{equation}
All the powers ${\bf tr} \left[  {\cal Y}^n  {\cal N} \right]$ can be
obtained by
differentiating over the phase angle parameter $\theta_Y$. Since the above
parameter-dependent invariant is robust to interactions between the
fermions, it
can be calculated for the noninteracting particles.
In the latter case the matrix ${\cal N}$ is diagonal with the eigenvalues $+1$
and $-1$ for right and left fermions correspondingly.  The trace of this
matrix
${\cal N}$ over given irreducible fermionic representation of the gauge group
is (with minus sign) the symbol $N_{(y/2,\underline a,I_W)}$ introduced by
Froggatt and Nielsen in Ref.\cite{FroggattNielsen}.  In their notations
$y/2(=Y)$ , $\underline a$ , and $I_W$ denote hypercharge, colour
representation and the weak isospin correspondingly.

 For the Standard  model
with right neutrino included one has
\begin{equation}
{\bf tr} ~ \left[ e^{i\theta_Y {\cal Y} } {\cal N} \right] = - {1 -\cos
\theta_Y/2\over 1 +\cos \theta_Y/2}
~ {\bf tr}   ~ \left[e^{i\theta_Y {\cal Y}}\right]~.
\label{GeneratingFunctionY}
\end{equation}
Here we used the values of the hypercharge: $Y=1/6$ for left quarks;
$Y=2/3$ and
$Y=-1/3$ for right quarks; $Y=-1/2$ for left leptons; $Y=-1$ for right
electron, muon, $\tau$ lepton, etc.; and
$Y=0$ for right neutrinos.

In addition to the hypercharge the weak charge is also conserved in the
Standard model above the electroweak transition. The generating function
for the
topological invariants which contain the powers of
both the hypercharge $Y$ and the weak charge $W_3$  has the form
\begin{equation}
{\bf tr} ~ \left[ e^{i\theta_W {\cal W}_3 } e^{i\theta_Y
{\cal Y}  }{\cal N}
\right]~  = {\cos
\theta_Y/2 -\cos
\theta_W/2\over \cos
\theta_Y/2 +\cos \theta_W/2}~{\bf tr}  ~ \left[e^{i\theta_W {\cal
W}_3}e^{i\theta_Y {\cal Y}  }\right].
\label{GeneratingFunctionWY}
\end{equation}
Finally one should add the powers of the generators ${\cal T}_{C3}$ and
${\cal T}_{C8}$ of the
$SU(3)$ colour group. They, however, do not change the  form of the generating
function in  Eq.(\ref{GeneratingFunctionWY}):
\begin{equation}
{ {\bf tr} ~ \left[ e^{i\theta_W
{\cal W}_3 } e^{i\theta_Y {\cal Y}} e^{i\theta_{C3} {\cal T}_{C3}
}e^{i\theta_{C8} {\cal T}_{C8} }~{\cal N}
\right] \over
{\bf tr}~\left[e^{i\theta_W {\cal W}_3}e^{i\theta_Y
{\cal Y}  }e^{i\theta_{C3} {\cal T}_{C3}}e^{i\theta_{C8} {\cal T}_{C8}} \right]
}~= {\cos
\theta_Y/2 -\cos
\theta_W/2\over \cos
\theta_Y/2 +\cos \theta_W/2} ~ .
\label{General}
\end{equation}
One could  add the powers of the generators of the groups
which correspond to conservation of the baryonic and leptonic charges. But
this also does not change the above result: the normalized topological
invariants
depend only on two angle-phase parameters $\theta_W$ and $\theta_Y$.

It is the
nonzero function  of the parameters determining the momentum-space invariants,
\begin{equation}
 {\cos
\theta_Y/2 -\cos
\theta_W/2\over \cos
\theta_Y/2 +\cos \theta_W/2} ~ ,
\label{FunctionOfParameters}
\end{equation}
that provides the mass protection for the Standard Model.

\section{Relation to Axial Anomaly}

The momentum-space topological invariants determine the
axial anomaly in fermionic systems. In the
2+1 condensed matter, the $\theta$-factor in front of the
Chern-Simons term is proportional to the invariant
Eq.(\ref{TopInvariantMatrix}), where the integration region $\sigma$ is the
whole 2+1 momentum space. For Quantum Hall Effect in electronic systems this
relation was established in
\cite{Ishikawa}. For $^3$He-A films, the Chern-Simons action for unit vector
$\hat d$  is \cite{VolovikYakovenko}
\begin{equation}
 S_\theta =  { \hbar \theta\over{32\pi^2}}\int
d^2x\hskip1mm dt\hskip1mm e^{\mu\nu\lambda}
  A_\mu   F_{\nu\lambda} ~,~F_{\nu\lambda}=\partial _\nu A_\lambda -
\partial_\lambda A_\nu=
\hat d\cdot\partial_\nu\left(\hat d\times\partial_\lambda\hat d\right)
 ~ ,
\label{ChernSimons}
\end{equation}
with
\begin{equation}
  \theta ={\pi\over 2}~{\bf tr} ~{\cal N}
 ~ .
\label{Theta}
\end{equation}
Here, in addition to spin indices, the matrix ${\cal N}$ contains the indices
of the transverse levels, which come from quantization of motion along
the normal to the film. The transverse levels play the role of different
families
of fermions.

The action in Eq.(\ref{ChernSimons}) represents the product of the
topological invariants in real and momentum spaces. For more general 2+1
condensed matter systems with different types of momentum-space invariants, see
\cite{Yakovenko}.

The Eq.(\ref{General}), which generates the momentum-space
invariants for the  Standard Model, is also related to anomalies. The
Wess-Zumino action, which describes the anomaly, can be also represented in
terms of the product of the real space and momentum space invariants. In
particular the axial anomaly term generated by the hypermagnetic field
$A^Y_\mu$ is
\begin{equation}
 S_{\rm WZ}^{Y}= {\hbar\over 96\pi^2}  \int
d^5x \hskip1mm e^{\mu\nu\lambda\alpha\beta}
  A^Y_\mu   F^Y_{\nu\lambda} F^Y_{\alpha\beta}~{\bf tr}  \left[   {\cal Y}^3
{\cal N} \right]
 ~.
\label{WessZumino}
\end{equation}
In the Standard model such action is zero because of the anomaly cancellation,
${\bf tr}\left[   {\cal Y}^3  {\cal N} \right]=0$, which follows from the
Eq.(\ref{General}). The other Wess-Zumino terms are also zero, again
due to anomaly cancellation expressions which follow from the
Eq.(\ref{General}):
\begin{equation}
{\bf tr}  \left[   {\cal Y}  {\cal N} \right] ={\bf tr}  \left[   {\cal Y}^3
{\cal N} \right]  = {\bf tr}  \left[   {\cal Y} {\cal
W}_3^2 {\cal N}
\right] =  {\bf tr}  \left[   {\cal Y}^2 {\cal
W}_3 {\cal N}
\right] =...=0~.
\label{AnomalyCancellation}
\end{equation}

\section{Masses via Symmetry Breaking}

The mass of the fermions can appear only if the $U(1)_Y\times SU(2)$
symmetry is
violated. Then the hypercharge and weak charge are no more conserved quantity.
The symmetry breaking which removes the mass protection can be for example the
complete symmetry breaking,
$U(1)_Y\times SU(2) \rightarrow 1$. In this case the only relevant
topological invariant which is left below the transition is
Eq.(\ref{TopInvariantMatrix}), but it is zero for the discussed model. Thus
there is no mass protection below such transition: all the fermions must
have the
mass.

There is another possibility, which
also removes the mass protection. It is the partial symmetry breaking
$U(1)_Y\times SU(2)
\rightarrow U(1)_Q$, where the generator $Q$ can be either $Q=\pm (Y-W_3)$ or
$Q=\pm (Y+W_3)$. As follows from Eq.(\ref{GeneratingFunctionWY}), the
generating
function for the topological invariants which contain the powers of $Q$ is
zero.
For example, for $Q=Y+W_3$, which corresponds to the electric charge, one has
\begin{equation}
{\bf tr} ~ \left[ e^{i\theta_Q {\cal Q} } {\cal N} \right] ={\bf tr} ~ \left[
e^{i\theta_Q {\cal Y} }  e^{i\theta_Q {\cal W}_3 }{\cal N} \right] =
0~,
\label{GeneratingFunctionQ}
\end{equation}
since this is the Eq.(\ref{GeneratingFunctionWY}) with $\theta_Y= \theta_W$.

The nature had chosen the electric charge to be $Q=Y+W_3$ and thus  each
elementary fermion  in our world has a mass.

\section{Discussion}

 What is the reason for such a choice? Why the nature had not
chosen the more natural symmetry breaking, such as
$U(1)_Y\times SU(2)
\rightarrow U(1)_Y$, $U(1)_Y\times SU(2)
\rightarrow SU(2)$ or $U(1)_Y\times SU(2)
\rightarrow U(1)_Y\times U(1)_{W_3}$? Or why it did not consider the symmetry
breaking
$U(1)_Y\times SU(2)
\rightarrow U(1)_Q$, with $Q=nY+mW_3$ and $n\neq \pm m$?
Probably this is
because in all the above cases the mass protection remains even below the
transition, and all fermions remain massless. This can shed light on the origin
of the electroweak transition. Maybe the elimination of the mass protection is
the only purpose of the transition.  This is similar to the Peierls
transition in
condensed matter: the formation of mass (gap) is not the consequence but the
cause of the transition. It might be energetically favourable to have
masses of
quasiparticles, since this leads to decrease of the energy of the Fermi sea.
Formation of the condensate of top quarks, which generates the
heavy mass of the top quark, could be a relevant scenario (see
review \cite{TaitPhD}).

Another question, why the hypercharges are so nicely organized that the
nullification of all the invariants ${\bf tr} ~ \left[  {\cal Q}^n
{\cal N} \right]$ is realized? The possible answer is that the group
$SU(3)\times SU(2) \times U(1)_Y$ is embedded in the higher symmetry group,
which contains the left-right symmetric operator $Q$. For example, it can be
the left-right symmetric model  of  Pati-Salam type
\cite{Foot} with symmetry group $SU(4)\times SU(2)_{L}\times SU(2)_{R}$.
It organizes  16 fermions
of each generation into the left and right baryon-lepton octets:
\begin{equation}
\matrix{ ~&SU(2)_L & SU(2)_R \cr
SU(4)&\left(\matrix{
u_L&d_L\cr
u_L&d_L\cr
u_L&d_L\cr
\nu_L&e_L\cr
}\right)
&
\left(\matrix{
u_R&d_R\cr
u_R&d_R\cr
u_R&d_R\cr
\nu_R&e_R\cr
}\right)\cr
}
\label{SU4}
\end{equation}

In addition to the $SU(3)$ charges there are the
right and left weak charges
$W_{R3}$ and $W_{RL}$, and the difference between the
baryonic and leptonic numbers $B-L$, which comes from the extension of the
colour group
$SU(3)$ to $SU(4)$:
\begin{equation}
\matrix{
Fermion&W_{L3}&W_{R3}&B-L\cr
u_L(3)&+{1\over 2}&0 &{1\over 3}\cr
d_L(3) & -{1\over 2}&0 &{1\over 3}\cr
u_R(3)&0 &+{1\over 2}  &{1\over 3}\cr
d_R(3) &0 &-{1\over 2}  &{1\over 3}\cr
e_L &-{1\over 2} &0  &-1\cr
\nu_L &+{1\over 2} &0  &-1\cr
e_R &0 &-{1\over 2}  &-1\cr
\nu_R &0 &+{1\over 2}  &-1\cr
}
\label{SU42}
\end{equation}
The generating function relevant for this group is
\begin{equation}
 {\bf tr}  \left[ e^{i\theta_R
{\cal W}^R_{3} }  e^{i\theta_L
{\cal W}^L_{3} } e^{i\theta_{BL} ({\cal B-L})}  ~{\cal N}
\right] = 2\left(\cos
{\theta_R\over 2} -\cos
{\theta_L\over 2}\right) \left( 3e^{i \theta_{BL}/3} +   e^{-i
\theta_{BL}}\right)  .
\label{SU43}
\end{equation}

The 16 fermions of one generation can be represented as the product $Cw$ of 4
bosons and 4 fermions \cite{Terazawa}. This scheme is similar to the
slave-boson
approach in condensed matter, where the particle is considered as a product of
the spinon and holon. Spinons are fermions which carry spin, while holons are
"slave"-bosons which carry electric charge \cite{Marchetti}.  In the Terazawa
scheme the "holons"
$C$ form the
$SU(4)$ quartet of spin-0
$SU(2)$-singlet particles which carry baryonic and leptonic charges, their
$B-L$ charges  of the
$SU(4)$ group are $( {1\over 3},  {1\over 3}, {1\over 3}, -1)$.  The "spinons"
are spin-${1\over 2}$ particles
$w$, which are
$SU(4)$ singlets and $SU(2)$-isodoublets; they carry spin and isospin.
\begin{equation}
\left(\matrix{
u_L&d_L&u_R&d_R\cr
u_L&d_L&u_R&d_R\cr
u_L&d_L&u_R&d_R\cr
\nu_L&e_L&\nu_R&e_R\cr
}\right)=\left(\matrix{
C_{1/3}\cr
C_{1/3}\cr
C_{1/3}\cr
C_{-1}\cr
}\right)\times
\left(\matrix{w_L^{+1/2}&
w_L^{-1/2}&w_R^{+1/2}&w_R^{-1/2}\cr}\right)
\label{Terazawa}
\end{equation}

Here $\pm 1/2$  means the charge $W_{L3}$ for the left spinons and  $W_{R3}$
for the right spinons, which coincides with the  electric charge of
spinons: $Q={1\over 2}(B-L) +W_{L3}+W_{R3}=W_{L3}+W_{R3}$. In Terazawa
notations
$w_1=(w_L^{+1/2},w_R^{+1/2})$ forms the doublet of spinons with $Q=+1/2$ and
$w_2=(w_L^{-1/2},w_R^{-1/2})$ --  with $Q=-1/2$.
These 4 spinons, 2 left and 2 right,
transform under
$SU(2)_{L}
\times SU(2)_{R}$ symmetry group. The generating function
for the momentum space  topological invariants for spinons is
${\bf tr}  \left[ e^{i\theta_R
{\cal W}^R_{3} }  e^{i\theta_L
{\cal W}^L_{3} }  ~{\cal N}\right]=2(\cos {\theta_R\over
2}-
\cos {\theta_L\over 2})$. The factorization in the Eq.(\ref{SU43}) thus
reflects the factorization in Eq.(\ref{Terazawa}).

The  electric charge $Q={1\over 2}(B-L) +W_{L3}+W_{R3}$ in the above scheme is
left-right symmetric. That is why, if only the electric charge is
conserved in the final broken symmetry state, the only relevant topological
invariant ${\bf tr} ~ \left[ e^{i\theta_Q {\cal Q} } {\cal N} \right]$ becomes
zero and the Weyl fermions can be paired into Dirac fermions. This fact
does not
depend on the definition of the hypercharge, which appears at the intermediate
stage where the symmetry is
$SU(3)\times SU(2)
\times U(1)_Y$. It also does not depend much on the definition of the
electric charge $Q$ itself: the only condition for the nullification of the
topological invariant is the symmetry (or antisymmetry) of $Q$ with respect to
the parity transformation.

%
%

\section*{ACKNOWLEDGMENTS}

 This work  was supported in part by the Russian
Foundation for Fundamental Research grant No. 96-02-16072 and by European
Science Foundation.

\end{document}